\documentclass[aps,prd,preprint,nofootinbib]{revtex4}
\usepackage{amsmath,amssymb,graphics,graphicx,epsfig,rotating,dcolumn}
\newcommand{\newc}{\newcommand}

\newc{\bsym}{\boldsymbol}
\newc{\mrm}{\mathrm}
\newc{\ovl}{\overline}
\newc{\ovla}{\overleftarrow}
\newc{\ovra}{\overrightarrow}
\newc{\wtil}{\widetilde}
\newc{\eps}{\epsilon}
\newc{\tri}{\triangle}
\newc{\hc}{\dagger}
\newc{\pd}{\partial}
\newc{\ra}{\rightarrow}
\newc{\lra}{\leftrightarrow}
\newc{\Lag}{{\mathcal{L}}}
\newc{\Dsl}{D\!\!\!/}
\newc{\half}{\frac{1}{2}}
\newc{\mc}{\multicolumn}
\newcolumntype{.}{D{.}{.}{-1}}

\begin{document}
\title{Twisted Mass Finite Volume Effects}
\author{Gilberto Colangelo}\email{gilberto@itp.unibe.ch}
\author{Urs Wenger}
\email{wenger@itp.unibe.ch}
\author{Jackson M. S. Wu}
\email{jbnwu@itp.unibe.ch}
\affiliation{Albert Einstein Center for Fundamental Physics\\
Institute for Theoretical Physics, University of Bern, Sidlerstrasse 5,
3012 Bern, Switzerland} 

\date{\today}

\begin{abstract}
We calculate finite volume effects on the pion masses and decay constant in
twisted mass lattice QCD (tmLQCD) at finite lattice spacing. We show that
the lighter neutral pion in tmLQCD gives rise to finite volume effects that
are exponentially enhanced when compared to those arising from the heavier 
charged pions. We demonstrate that the recent two flavour twisted mass 
lattice data can be better fitted when twisted mass effects in finite 
volume corrections are taken into account.
\end{abstract}


\maketitle

\section{Introduction}
For large enough lattices, finite volume effects (FVEs) are only sensitive
to the long distance physics of the underlying theory -- for lattice QCD
simulations they are well described by chiral perturbation theory
(ChPT)~\cite{Gasser:1987zq}, and can be analyzed systematically in a chiral
expansion. In a series of papers~\cite{CD04,CH04,CDH05}, FVEs on
pseudoscalar meson masses and decay constants in the continuum limit have
been studied with the help of resummed L\"uscher formulae used in
conjunction with ChPT. In the original asymptotic formula by
L\"uscher~\cite{Lus86}, the exponentially dominating FVEs for the mass of a
particle $P$ is expressed as an integral over the forward scattering
amplitude of the particle $P$ off the lightest particle in the spectrum
(pions for QCD).  Since the scattering amplitude is needed only at low
energy, it can be evaluated in ChPT. In Ref.~\cite{CDH05}, it has been
shown that by resumming a series of such integrals (over the same
scattering amplitude but with kernels increasingly exponentially
suppressed), one can improve the reliability of such formulae in describing
FVEs.

As shown in Ref.~\cite{CDH05}, if one inserts the tree-level scattering
amplitude in the resummed L\"uscher formula, one obtains exactly the
one-loop ChPT calculation of the FVEs. In Ref~\cite{CH06}, the pion mass
has been calculated in ChPT to two loops and compared with the result
obtained from the resummed L\"uscher formula using the one-loop ChPT
representation of the scattering amplitude. The difference is found to be
very small, and one has thus confidence in the validity of the resummed
L\"uscher formulae in predicting the size of FVEs. This allows for a much
simpler evaluation of the main effects at the two-loop level and beyond,
and for one to check the convergence of the chiral expansion for FVEs.
Nevertheless, it is still essential to test these predictions against
actual data.

Recently, the European Twisted Mass Collaboration (ETMC) has provided one
such test for the pion mass and decay constant~\cite{Urb07,ETMC09}, and the
outcome was not entirely positive. While FVEs on the decay constant are
rather well described by the resummed formula at the next-to-leading order
(NLO), the same does not hold for the pion mass. For the pion mass, FVEs
are larger than what is predicted from the NLO and NNLO resummed formulae
(see Table 2 in Ref.~\cite{Urb07}), and the discrepancy appears to be much
larger than the size of the error estimated in Ref.~\cite{CDH05}, which
calls for an explanation.

Before discussing possible sources of such a deviation, we stress that the
ETMC data are on the borderline of applicability of ChPT for calculating
finite volume corrections. It has been argued in Ref.~\cite{CD04} that for
ChPT to be applicable, a box size of $L > 2$~fm is necessary. The
comparison made in Table 2 of Ref.~\cite{Urb07} uses the largest volumes as
reference points, and FVEs are measured with respect to these. The data
sets used to study FVEs thus have $L \sim 2.0$~fm (where ChPT may still
marginally work) or less (where ChPT is not expected to be valid). This may 
be a possible reason for the discrepancy.

In this paper we study another possible source for deviation from the
continuum formulae, i.e. discretization effects. As is well known, at
finite lattice spacing, isospin- and parity-breaking effects are sizeable
in twisted mass lattice QCD (tmLQCD). In particular, the neutral pion mass
becomes smaller than the charged ones~\footnote{For the range of masses and
  lattice spacings of the ETMC simulations, the neutral pion is 
  typically about 15-20\% lighter than the charged ones.}, 
and parity-breaking interactions among pions become possible. Both of these
effects are well described in the framework of
tmChPT~\cite{MS04,Sco04,SW04,SW05}, and they have nontrivial influences on
the finite volume corrections as they generate exponentially enhanced
FVEs. Indeed, if one takes the continuum limit first without explicitly
accounting for these effects and make finite volume corrections using the
continuum formulae, one will not be able to fully disentangle in the final
results FVEs from discretization effects.\footnote{The size of the twisted
  mass discretization effects can only be determined after a detailed
  analysis. Nevertheless, one can argue that if these were responsible for
  the discrepancy between the FVEs observed by ETMC and those
  calculated in the continuum, they are at the percent level.}

The rest of the paper is organized as follows. In the next section, we
briefly describe the formalism of tmChPT, highlighting aspects that are
relevant for our calculations. In Sec.~\ref{sec:tmFS}, we explain why and
which discretization effects are exponentially enhanced at finite volume
with the help of LO asymptotic formulae. In Sec.~\ref{sec:RAF}, we give
resummed asymptotic formulae at NLO applicable to the ETMC data, which we use
to perform a new analysis on the ETMC data. Sec.~\ref{sec:NA} contains relevant
results and discussion of our analysis, and Sec.~\ref{sec:concl} our conclusion.
We concentrate in this paper on the charged pions where high quality lattice
data are available, but also provide formulae for FVE for the neutral pion
mass for future use.

\section{\label{sec:tmChPT} Twisted mass chiral effective theory}
Consider tmLQCD with a degenerate doublet of quarks~\cite{FR04}. The low
energy, long distance dynamics of the underlying lattice theory can be
described by an effective continuum chiral theory constructed using the
two-step procedure of Ref.~\cite{SS98}, whereby effects of discretization
errors are systematically incorporated in a joint expansion of the lattice
spacing, $a$, the quark mass, $m$, and the twisted mass, $\mu$. This was
carried out to NLO in Ref.~\cite{SW04}, and the resulting tmChPT studied in
detail in Ref.~\cite{SW05}. Using the power counting scheme, $m \sim \mu
\sim p^2 \sim a\Lambda_{\rm QCD}^2$, the twisted mass effective chiral
Lagrangian reads~\cite{SW05}:
\begin{align}\label{eq:tmLCh}
\mathcal{L}_\chi &=  
\frac{F^2}{4} \langle D_\mu U D_\mu U^\hc -( \chi'^{\hc}U +U^\hc\chi') \rangle \notag \\ 
&\quad
- \frac{\ell_1}{4} \langle D_\mu U D_\mu U^\hc\rangle^2 
- \frac{\ell_2}{4} \langle D_\mu U D_\nu U^\hc \rangle^2
- \frac{\ell_3}{16} \langle \chi'^{\hc} U + U^\hc \chi' \rangle^2
\notag \\
&\quad 
+ \frac{\ell_4}{4} \langle D_\mu \chi'^\hc D_\mu U + D_\mu U^\hc D_\mu \chi'\rangle \notag \\ 
&\quad
- W \langle \chi'^{\hc} U + U^\hc\chi' \rangle \langle \hat{A}^{\hc} U + U^{\hc}\hat{A} \rangle
- W'\langle\hat{A}^{\hc}U + U^{\hc}\hat{A}\rangle^2 \notag \\
&\quad
+ W_{10} \langle D_\mu\hat{A}^\hc D_\mu U + D_\mu U^\hc D_\mu\hat{A} \rangle
+ \wtil{W} \langle D_\mu U D_\mu U^\hc \rangle \langle \hat{A}^{\hc} U + U^{\hc}\hat{A} \rangle
\notag \\
&\quad
- H' \langle \hat{A}^\hc\chi'+ \chi'^\hc\hat{A}\rangle \,,
\end{align}
where we have displayed only parts relevant for our discussion and our
study of FVEs below. Here, $F$ is normalized so that $F_\pi = 92.4$~MeV,
$U$ is the usual $SU(2)$ matrix-valued field, $\langle\ldots\rangle$
denotes the trace, $\ell_i$ the usual Gasser-Leutwyler Low Energy Constants
(LECs), and the rest of coefficients LECs arising from discretization
errors. The quantities $\chi'$ and $\hat{A}$ are spurions for quark masses
and discretization errors that are set to constant values at the end of the
analysis:
\begin{equation}
\chi' \ra 2B_0(m + i\tau_3\mu) + 2 W_0\,a \equiv \hat{m} + \hat{a} +
i\tau_3\hat{\mu} \,, \qquad \hat{A} \ra 2 W_0\,a \equiv \hat{a} \,, 
\end{equation}
where $\tau_3$ is normalized so that $\tau_3^2 = 1$, $B_0$ and $W_0$ are
unknown dimensionful constants, and we have defined the quantities
$\hat{m}$, $\hat{\mu}$, and $\hat{a}$.

As explained in Ref.~\cite{SW05}, the $W_{10}$ term in the effective chiral
Lagrangian is redundant, and can be transformed away into a combination of
$W$, $\wtil{W}$ and $H'$ terms. However, the same redundancy can be
used to eliminate the $\wtil{W}$ term in favour of the $W_{10}$ term
instead, which has the advantage of simplifying the Feynman rules when
studying FVEs in tmChPT, as we see below. In doing so, results derived in
Ref.~\cite{SW05} would remain the same except with all $\wtil{W}$ terms
removed. In particular, the expansion about the NLO vacuum with external
fields set to zero now
reads: 
\begin{equation}\label{eq:feynman}
\mathcal{L}_\chi = 
\half\pd_\mu\vec\pi\cdot\pd_\mu\vec\pi
+\frac{\pi^2}{2}(M' + \Delta M') - \frac{\pi_3^2}{2}\frac{32}{F^2}\hat{a}^2
s^2 W' 
+\frac{\pi_3\pi^2}{2}\frac{\epsilon M'}{F} 
-(\pi^2)^2\frac{M'}{24F^2} + \dots \,, 
\end{equation}
where $\pi^2 = \vec\pi\cdot\vec\pi$, and we have used the convenient
parametrization $U = \sigma + i\vec{\pi}\cdot\vec{\tau}/F$, with $\sigma =
\sqrt{1-\pi^2/F^2}$. 
The quantities $s$ and $c$ denote the sine and cosine of a nonperturbatively 
defined vacuum (twist) angle $\omega$, and $\eps$ is the shift of the vacuum 
angle at NLO tmChPT from that at LO:
\begin{equation}\label{eq:GSMeps}
\epsilon \equiv \omega_m - \omega_0
= -\frac{16}{F^2}\hat{a}s_0\left(W + 2W'\hat{a}c_0/M'\right) \,.
\end{equation}
Note that $\omega$ differs from both $\omega_0$ and $\omega_m$ by
$\mathcal{O}(a)$ so that at NLO accuracy, $s$ and $c$ could equally well be
$s_0$ ($s_m$) and $c_0$ ($c_m$). The mass parameters are given by
\begin{equation}
M' = |\chi'| = \sqrt{(\hat{m} + \hat{a})^2 + \hat{\mu}^2} \,, \qquad
\Delta M' = \frac{2}{F^2}\left[M'^2 \ell_3 + 16( \hat{a}c M' W + \hat{a}^2 c^2 W' ) \right] \,,
\end{equation}
and the charged and neutral pion masses at NLO read:
\begin{align}\label{eq:mpi2}
M_{\pi^\pm}^2 &= 
M'\left[ 1 - \frac{M'}{32 \pi^2 F^2} \bar \ell_3 
+ \frac{32 M'}{F^2}\hat{a}c (M' W +  \hat{a} c W') \right]  \,, \\
M_{\pi^0}^2 &= M_{\pi^\pm}^2 - \frac{32}{F^2}\hat{a}^2 s^2 W' \equiv
M_{\pi^\pm}^2 - a^2 K \,, \quad 
K = \frac{128}{F^2}s^2 W_0^2 W' \,,
\end{align}
with 
\begin{equation}
\bar{\ell}_i = \bar{\ell}_i^{\,phys} + 2\log\frac{M_\pi^{phys}}{M_\pi}
\equiv \log\frac{\Lambda_i^2}{M_\pi^2} 
\end{equation}
the standard scale-independent LECs of $SU(2)$ ChPT~\cite{Colangelo:2001df}. We have
defined the dimensionful quantities $\Lambda_i$ for use in our numerical
analysis below. The pion decay constant at NLO reads 
\begin{equation}\label{eq:Fpi}
F_\pi \equiv F_{\pi^\pm} =
F\left[1+\frac{M'}{16 \pi^2 F^2} \bar\ell_4 + \frac{4 M'}{F^2} \hat{a}c W_{10}\right] \,.
\end{equation}
Note that at maximal twist, $\omega = \pi/2$, the charged pion masses and
pion decay constant above take their continuum form with $\mathcal{O}(a)$
effects automatically removed. At NLO, the effects of twisting reside only
in the $\mathcal{O}(a^2)$ charged-neutral pion mass splitting, which is
maximal at maximal twist.

\section{\label{sec:tmFS} Exponentially enhanced twisted mass discretization effects at finite volume}
Consider L\"uscher's formula for the pion mass~\footnote{For clarity in
  demonstrating the idea here, we do not write out the resummed version of
  this formula, but will defer to the next section where we give our final 
formulae in full.}:
\begin{equation}\label{eq:mpiluscher}
\frac{M_\pi(L) - M_\pi}{M_\pi} = -\frac{3}{16\pi^2\lambda_\pi}
\int\limits_{-\infty}^{\infty}\!\!dy\,\mathcal{F}(iy)e^{-\sqrt{1+y^2}\lambda_\pi}
+ \ldots \,,   
\qquad \lambda_\pi \equiv M_\pi L \,,
\end{equation}
where $\mathcal{F}(iy)$ is the forward $\pi\pi$-scattering amplitude
evaluated for a purely imaginary argument (given in units of pion mass),
and $y$ is real and dimensionless. This formula gives the exponentially
dominant contribution under the assumption that the pion is the lightest
particle in the spectrum. In the presence of a splitting between the
charged and the neutral pions, as is the case in tmLQCD, this formula is
modified to
\begin{equation}
R_i\equiv\frac{M_{\pi^i}(L) - M_{\pi^i}}{M_{\pi^i}} =
-\frac{3}{16\pi^2\lambda_{i}}\sum_{j=1}^3\frac{M_{\pi^j}}{M_{\pi^i}}
\int\limits_{-\infty}^{\infty}\!\!dy\,\mathcal{F}_{ij}(iy)e^{-\sqrt{1+y^2}\lambda_j} + \ldots \,,
\end{equation}
where $\mathcal{F}_{ij}(iy)$ is now the forward scattering amplitude of
pions with isospin index $i$ off pions with isospin index $j$. Note that
$y$ is still dimensionless but now normalized to $M_{\pi^j}$, and we use
the shorthand $\lambda_j$ for $\lambda_{\pi^j}$.

With its lighter mass, the neutral pion contribution is (exponentially)
dominant with respect to the heavier charged pions. The weights of these
contributions are given by the forward scattering amplitudes, and at LO in
tmChPT they read:
\begin{equation}\label{eq:FLO}
\mathcal{F}_{11} = -\mathcal{F}_{12} = -\mathcal{F}_{13} = \frac{M'}{F^2} \,,
\end{equation}
where $M'$ and $F$ are the LO pion mass-squared and decay constant. This
shows that since pions with index 1 and 2 are degenerate, their
contributions cancel. The relative finite-volume shift for the charged pion
mass at LO in tmChPT is then given by
\begin{equation}
R_\pm = \frac{3}{8\pi^2\lambda_\pm}\frac{M_{\pi^0}}{M_{\pi^\pm}}\frac{M'}{F^2}K_1(\lambda_{0}) 
+ \ldots \,,
\end{equation}
where $K_i$ denotes the modified Bessel function. As an illustration of
the importance of this discretization effect, we evaluate $R_\pm$ for the
ETMC $B_1$ ensemble in Ref.~\cite{Urb07}, for which $M_{\pi^\pm} =
0.33$~GeV and $M_{\pi^0} = 0.27$ GeV:
\begin{equation}
R_\pm = 0.26\% \qquad \mbox{(B1 ensemble)} \,.
\end{equation}
However, if we set $M_{\pi^0} = M_{\pi^\pm} = 0.33$~GeV, i.e. turning off
the twisted mass effects, we get 
\begin{equation}
R_\pm = 0.15\% \qquad \mbox{(B1 ensemble)} \,.
\end{equation}
We note that if one here resums the whole series of exponentially
subdominant LO contributions (corresponding to a LO ChPT evaluation of
FVEs), 
0.15\% goes up to 0.62\%; using the NNLO resummed asymptotic formula of
Ref.~\cite{CDH05} gives 1\%, whereas the ETMC measurement is 1.8(5)\%. 

There is a second discretization effect arising from the parity-violating
cubic interactions (see Eq.~\eqref{eq:feynman}), which could be potentially
significant as well. The contributions to FVEs due to these in the pion
mass formally come in at higher order: they are NNLO FVEs, or
$\mathcal{O}(p^8)$ corrections to the pion mass.\footnote{In the tmChPT
  counting we use, these vertices are $\mathcal{O}(p^4)$, and one needs at
  least two of them to contribute to the $\pi\pi$-scattering amplitude
  entering L\"uscher's formula given by
  Eq.~\eqref{eq:mpiluscher}. Equivalently, one needs at least two such
  vertices to make a self-energy correction to the pion propagator that
  yields FVEs.}  But because of the different topology of the relevant loop
diagrams, these contributions are exponentially enhanced with respect to
the tadpole diagrams that contain virtual neutral pions. For these
diagrams, the dominating exponential behavior goes as
$e^{-\lambda_0\sqrt{1-w^2}}$, where $w \equiv M_{\pi^0}/(2M_{\pi^\pm})$.
From the parameters extracted from the $B_1$ ensemble, the enhancement of
such a contribution with respect to the standard $e^{-\lambda_\pm}$
behavior is more than 200\%, i.e.
$e^{-\lambda_0\sqrt{1-w^2}}/e^{-\lambda_\pm} \simeq 2.4$, which motivates
us to investigate these effects in our analysis even when they are formally
of higher order.  Now these contributions to the FVEs are proportional to
$\epsilon^2$, which involves unknown LECs $W$ and $W'$ that have to be
determined if the size of these contributions are to be predicted. If one
works at maximal twist, only $W$ is required, which one can determine
e.g. from the ratio of density matrix elements~\cite{SW05}.\footnote{The
  LEC $W'$ can be determined from the mass splitting between the charged
  and neutral pion~\cite{SW05}. But this is difficult in practice as
  calculating the neutral pion mass on the lattice involves quark
  disconnected contributions.} We remark here that if the exponentially
enhanced parity-violating contributions are truly large or that FVEs can be
measured precise enough, one can turn it around and use instead FVEs to get
a measure of $W$; one could even determine $W_0$, the additional unknown
dimensionful constant in tmChPT, if $W$ is already determined elsewhere.

\section{\label{sec:RAF} NLO formulae for the pion masses and decay constant
  in finite volume} 
\subsection{The standard contributions}
We provide here the complete resummed asymptotic formulae at NLO for the
relative finite volume shift of the charged pion mass and decay
constant. We split the contributions to the NLO FVEs for the charged pion
mass into contributions in decreasing exponential importance:
\begin{equation}\label{eq:RM}
R_{M_\pm} =  R_{M_\pm}(\lambda_0) + R_{M_\pm}(\lambda_\pm) \,,
\end{equation}
where $R_{M_\pm}(\lambda_0)$ is the standard contribution due to neutral
pions traveling around the whole volume ($\sim e^{-\lambda_0}$), and
$R_{M_\pm}(\lambda_\pm)$ that due to the charged pions ($\sim
e^{-\lambda_\pm}$).

The standard contributions $R_{M_\pm}$ start at $\mathcal{O}(p^2)$, and it
is easy to obtain the next order correction by evaluating
$\pi\pi$-scattering at NLO in tmChPT.\footnote{For simplicity, we have set
  $M_{\pi^0}=M_{\pi^\pm}$ in the NLO $\pi\pi$-scattering amplitude, as the
  effects of the charged-neutral pion mass splitting is higher order. The
  same goes for the amplitude entering the L\"uscher formula for $F_\pi$.}
Inserting this into the resummed L\"uscher formula, we have:
\begin{align}\label{eq:RMNLO}
R_{M_\pm}(\lambda_i) &=
-\frac{\xi_\pm}{2\lambda_\pm} \sum_{n=1}^{\infty}
\frac{m(n)}{\sqrt{n}}\left[  
I^{(2)}_{M,i}(\sqrt{n}\lambda_i) + \xi_\pm I^{(4)}_{M,i}(\sqrt{n}\lambda_i) +
\mathcal{O}(\xi_\pm^2)\right] \,, \quad 
\xi_\pm \equiv \frac{M_{\pi^\pm}^2}{16\pi^2 F_\pi^2} \,,
\end{align}
where $i\in\{\pm,\,0\}$ is the isospin index, 
$m(n)$ is the multiplicity of the integer vector $\vec{n}$ of length $n = |\vec{n}|$, and 
\begin{align}
I^{(2)}_{M,0} &= -B^0 \,, \qquad I^{(2)}_{M,\pm} = 0 \,, \\
I^{(4)}_{M,0} &= \left[\frac{4}{3}\bar{\ell}_1 - \frac{1}{2}\bar{\ell}_3
    - 2\bar{\ell}_4 + \frac{13}{18}\right]\!\!B^0 + \left[\frac{20}{9} -
    \frac{8}{3}\bar{\ell}_2\right]\!\!  B^2 +
  \frac{2}{3}\!\left(R^0_0 + 2R^1_0 - 4R^2_0\right) \,, \notag \\ 
I^{(4)}_{M,\pm} &= \left[\frac{8}{3}(\bar{\ell}_1 + \bar{\ell}_2) -
    2\bar{\ell}_3 - \frac{34}{9}\right]\!\!B^0 +
  \left[\frac{92}{9}-\frac{8}{3}\bar{\ell}_1 - 8\bar{\ell}_2\right]\!\!B^2 +
  \frac{1}{3}\!\left(11R^0_0 - 20R^1_0 - 32R^2_0\right) \,, \notag
\end{align}
with $B^{2k} \equiv B^{2k}(\sqrt{n}\lambda_i)$ and 
$R^{k\,(\prime)}_0 \equiv R^{k\,(\prime)}_0(\sqrt{n}\lambda_i)$ integrals given by
\begin{align}
B^{2k}(\sqrt{n}\lambda_i) &= r_i^{2k+1}\!\!\int\limits_{-\infty}^{\infty}\!\!dy\,y^{2k}e^{-\sqrt{n(1+y^2)}\lambda_i} = 
r_i^{2k+1}\frac{\Gamma(k+1/2)}{\Gamma(3/2)}\left(\frac{2}{\sqrt{n}\lambda_i}\right)^k
K_{k+1}(\sqrt{n}\lambda_i) \,, \\
R^{k\,(\prime)}_0(\sqrt{n}\lambda_i) &= \bigg\{\genfrac{}{}{0pt}{}{\mrm{Re}}{\mrm{Im}}\,
\int\limits_{-\infty}^{\infty}\!\!dy\, y^k e^{-\sqrt{n[1+(y/r_i)^2]}\lambda_i}g^{(\prime)}(2 + 2iy) \,, 
\quad \mrm{for}\;\bigg\{\genfrac{}{}{0pt}{}{k\;\mrm{even}}{k\;\mrm{odd}} \,,
\end{align}
where $r_i=M_{\pi^i}/M_{\pi^\pm}$ and~\footnote{The function $g(x)$ is
  related to the standard $\bar J$ one-loop function through  
$g(x) = 16\pi^2 \bar{J}(x M_{\pi^\pm}^2)$.} 
\begin{equation}
g(x)=\sigma\log\frac{\sigma - 1}{\sigma + 1}+2 \,, \qquad \sigma(x) =
\sqrt{1-4/x} \,. 
\end{equation}

For completeness we give here also the full NLO resummed asymptotic formula
for FVEs in the neutral pion mass, although these will not be used in our
numerical analysis. The formula reads:
\begin{equation}
R_{M_0} =  R_{M_0}(\lambda_0) + R_{M_0}(\lambda_\pm) \,,
\label{eq:RM0}
\end{equation}
where 
\begin{align}
R_{M_0}(\lambda_i) &=
-\frac{\xi_\pm}{2\lambda_0} \sum_{n=1}^{\infty}
\frac{m(n)}{\sqrt{n}}\left[ I^{(2)}_{M_0,i}(\sqrt{n}\lambda_i) 
+ \xi_\pm I^{(4)}_{M_0,i}(\sqrt{n}\lambda_i) +
  \mathcal{O}(\xi_\pm^2)\right] \, .
\end{align}
The integrals $I^{(n)}_{M_0,i}$ can be expressed in terms of the
$I^{(n)}_{M,i}$ integrals defined for the charged pion mass:
\begin{eqnarray}
  I^{(n)}_{M_0,0}(\sqrt{n} \lambda_0)&=&I^{(n)}_{M,\pm}(\sqrt{n}
  \lambda_0)-I^{(n)}_{M,0}(\sqrt{n} \lambda_0) \nonumber \\ 
I^{(n)}_{M_0,\pm}(\sqrt{n} \lambda_\pm)&=&2 I^{(n)}_{M,\pm}(\sqrt{n}
\lambda_\pm) \; \; . 
\end{eqnarray}

For the pion decay constants, the decomposition of the finite volume shifts
is similar to that given for the pion mass. We have:
\begin{equation}\label{eq:RFNLO}
R_{F_\pm} = R_{F_{\pm}}(\lambda_0) + R_{F_{\pm}}(\lambda_\pm) \,, 
\end{equation}
where
\begin{align}
R_{F_{\pm}}(\lambda_i) &=
\frac{\xi_\pm}{\lambda_\pm} \sum_{n=1}^{\infty}
\frac{m(n)}{\sqrt{n}}\left[ I^{(2)}_{F,i}(\sqrt{n}\lambda_i) + \xi_\pm
  I^{(4)}_{F,i}(\sqrt{n}\lambda_i) + \mathcal{O}(\xi_\pm^2)\right] \,, 
\end{align}
and
\begin{align}
I^{(2)}_{F,0}  &= I^{(2)}_{F,\pm} = -B^0 \,, \notag \\
I^{(4)}_{F,0} &= 
\left[\frac{1}{9} + \frac{2}{3}\bar{\ell}_1 - \bar{\ell}_4\right]\!\!B^0 + 
\left[\frac{20}{9} - \frac{8}{3}\bar{\ell}_2\right]\!\!B^2 + 
\frac{1}{3}\!\left(2R^0_0 + 4R^1_0 - 8R^2_0 
-R^{0\,\prime}_0 - 2R^{1\,\prime}_0 + 4R^{2\,\prime}_0\right) \,, \notag \\ 
I^{(4)}_{F,\pm} &= 
\left[-\frac{8}{9} + \frac{4}{3}\bar{\ell}_1 + \frac{4}{3}\bar{\ell}_2 - 2\bar{\ell}_4\right]\!\!B^0 +  
\left[\frac{92}{9} - \frac{8}{3}\bar{\ell}_1 - 8\bar{\ell}_2\right]\!\!B^2 \notag \\
&\quad +\frac{1}{3}\!\left[2R^0_0 - 8R^1_0 - 32R^2_0 
-\frac{11}{2}R^{0\,\prime}_0 + 10R^{1\,\prime}_0 + 16R^{2\,\prime}_0\right] \,.
\end{align}

\subsection{The parity-violating cubic interaction contributions}
As discussed above, because of the exponential enhancement, contributions
due to parity-violating cubic interactions in tmChPT may have an effect at
NLO in addition to the standard contributions to FVEs (arising from
parity-conserving quartic interactions), despite being formally higher order.
For the charged pion mass, they are given by
\begin{equation}\label{eq:DRM}
\Delta R_{M_\pm} =
-\eps^2\frac{\xi_\pm}{\lambda_\pm}\sum_{n=1}^{\infty}\frac{m(n)}{\sqrt{n}} 
\left[\pi e^{-\lambda_0\sqrt{n(1-w^2)}} -
  \frac{1}{4}\int\limits_{-\infty}^{\infty}\!\!d y\, 
\frac{e^{-\lambda_0 \sqrt{n(1+ y^2)}}}{ y^2+w^2}\right] \,.
\end{equation}
Recall that $w \equiv M_{\pi^0}/(2M_{\pi^\pm}) < 1$. As explained above,
the leading exponential contributions to pion mass in finite volume
arises first at $\mathcal{O}(p^8)$ in the chiral counting. Thus,
contributions to the relative shift given here is $\mathcal{O}(p^6)$. Note
that the sign of $\Delta R_{M_\pm}$ is fixed, and is opposite to the
standard contributions given in Eq.~(\ref{eq:RM}), which tend to make the pion 
masses larger; $\Delta R_{M_\pm}$ would hence reduce this enlargement.

For the pion decay constants, the exponentially enhanced contributions
arise from the parity-violating interaction with the axial current. These
have the same form as that for the charged pion mass except for a
different prefactor involving the LEC $W_{10}$:
\begin{equation}\label{eq:DRFp}
\Delta R_{F_\pm} = -\frac{3s\,\delta}{4\eps}\Delta R_{M_\pm} \,, \quad 
\delta = \frac{4\hat{a}W_{10}}{F^2} \,.
\end{equation}
Note that $\Delta R_{F_\pm}$ is actually proportional to $\epsilon$, and
unlike the case for pion mass above, its sign is not fixed because
$W_{10}$ is unknown. We also remark that this contribution to the FVEs
depends on how exactly $F_\pi$ is calculated. If the pseudoscalar density
is used instead of the axial current as is the case here, $\Delta
R_{F_\pm}$ would be proportional to $W$ instead of $W_{10}$. This is a
consequence of the fact that the Ward identity
\begin{equation}
G_\pi m_q = F_\pi M_\pi^2
\end{equation}
is violated by discretization effects in tmChPT.

\section{\label{sec:NA} Numerical analysis}
As a first illustration of the importance of the full NLO discretization
effects in finite volume corrections, we evaluate them numerically using
the formulae derived in the previous section for a box of size 2~fm and
with pion masses close to those in the ETMC ensemble B1~\cite{Urb07,ETMC09}. The
results are shown in Table~\ref{tab:NLOeg}, where we give not only the LO 
and NLO standard contributions, but also the parity-violating cubic 
interaction contributions.
\begin{table}[t]
\begin{ruledtabular}
\begin{tabular}{c.....}
Relative shift & \mc{2}{c}{LO} & \mc{2}{c}{NLO} & \mc{1}{c}{Cubic interactions} \\
(\%)           & \mc{1}{c}{ChPT} & \mc{1}{c}{tmChPT} & \mc{1}{c}{ChPT} & \mc{1}{c}{tmChPT} & \\  
\hline
$M_\pi$ & 0.38 & 0.71 & 0.74 & 1.24 & -0.08 \\
$F_\pi$ & -1.5 & -2.2 & -2.1 & -3.0 & 0.06
\end{tabular}
\end{ruledtabular}
\caption{\label{tab:NLOeg}
Relative finite volume corrections in percentage for a $L = 2$~fm
box. The ChPT columns denote the case of degenerate pion with
$M_{\pi^\pm} = M_{\pi^0} = 0.33$~GeV, while the tmChPT columns that where 
charged and neutral pions are split with $M_{\pi^\pm}=0.33$~GeV and 
$M_{\pi^0}=0.27$~GeV.At these mass values, $F_\pi$ is about and so taken to
be 0.11~GeV \cite{CD04}. 
The $3\pi$-vertex contributions are obtained by 
setting $\eps=\xi$ and $s\delta/\eps=1$.} 
\end{table}
As discussed above, these involve unknown LECs. So to evaluate them, we set
$\epsilon = \xi$ and $s\delta/\epsilon = 1$, which is a conservative
choice.  Crucially, we see when compared to the standard contributions, those
arising from the parity-violating cubic interactions turn out to be subdominant 
after all, despite the exponential enhancement. Note as observed above, in 
the case of $M_\pi$ the two types of contributions have opposite signs so that 
finite volume corrections are reduced (albeit only slightly). In the $F_\pi$ case,
the sign is chosen when we set $s\delta/\epsilon = 1$; it is otherwise free
in general.

Overall, Table~\ref{tab:NLOeg} shows that the corrections to the continuum
formulae are substantial, expecially for the pion masses, and this argues
for a more detailed analysis of the ETMC data. Using the most recent ETMC
data~\cite{ETMC09}, we performed a new analysis to evaluate the impact on
the finite volume corrections when effects due to the lighter neutral pion
are included. A global $\chi^2$ fit to the ensembles
$B_1-B_4,B_6,B_7,C_1-C_3$ and $D_1$ is performed, which all have a charged
pion mass below 500~MeV and a box size of at least 2~fm (so that ChPT can
be safely applied).\footnote{Ensembles $C_5$ and $D_2$ have $L \simeq
  1.6$~fm, and thus too small a volume to have the FVEs reliably described
  by our formulae. Nevertheless, we have checked that including them in our
  fits does not deteriorate the quality of the fits, nor changes any of our
  conclusions.} Our fit is purely statistical and does not include systematic
errors, which have not been released. We recall that ensembles in the B (D)
set have a coarser (finer) lattice than the C set.

The forms we use to fit the ETMC data on the charged pion masses and decay
constants at maximal twist read:
\begin{align}
M_{\pi^\pm}(a,L) &= 
\sqrt{M'}\left[ 1 - \frac{M'}{32 \pi^2 F^2} \bar \ell_3 + a^2
  D_m\right]^{1/2}(1 + R_{M_\pm}) \,, \\ 
F_\pi(a,L) &= F\left[1+\frac{M'}{16 \pi^2 F^2} \bar\ell_4 + a^2
  D_f\right](1 + R_{F_\pm}) \,. 
\end{align}
Notice that we have not included $\Delta R_{M_\pm}$ and $\Delta R_{F_\pm}$
in our fitting forms: we have checked that the data at their current
precision, have no sensitivity to these higher order parity-violating
effects.\footnote{When fitting with $\Delta R_{M_\pm}$ and $\Delta
  R_{F_\pm}$ included, we find that $\epsilon$ is driven to be vanishingly
  small, as the data do not favour a reduction of FVEs in the charged
  pion masses that $\Delta R_{M_\pm}$ necessarily brings about. Because of
  this, fluctuations in the pion decay constant data tend to drive $\delta$ 
  into a run-away increase, as it has to compensate for the smallness of
  $\epsilon$ (see Eq.~\eqref{eq:DRFp}), and this is seen in the fitting. 
  As a result, no firm conclusions can be drawn about the presence of these
  parity-violating contributions from the data.} We have however included
parameters $D_m$ and $D_f$ in our fit, which account for the relative
$\mathcal{O}(a^2)$ effects in $M_\pi$ and $F_\pi$ respectively. Although
such effects are formally higher order (NNLO) in the counting we use, they
may appear spuriously if the maximal twist condition, $\omega = \pi/2$, is
only determined to $\mathcal{O}(a)$ accuracy. They are therefore included to
provide useful diagnostics. 

In all, our fitting parameters at maximal twist thus consist of the tmChPT
parameters (in lattice units) $a_C F$, $2a_C B_0$, $a_C\Lambda_3$,
$a_C\Lambda_4$ and $a_C^4 K$, the ratios of lattice spacings $a_B/a_C$ and
$a_D/a_C$, and additional lattice discretization parameters $a_C^2 D_m$ and
$a_C^2 D_f$. The notation $a_X$ here denotes the lattice spacing for the
ensemble set $X$, and we have chosen $a_C$ to be the reference lattice
unit. The scale setting is accomplished by fixing $F_\pi = 92.4$~MeV at the
point where the ratio $M_\pi/F_\pi$ assumes its physical value.  Note that
the $r_0/a_X$ data provide constraints on the ratios of lattice
spacings. In our fitting the values of $r_0/a_X$ are taken in the chiral limit
as provided in Ref.~\cite{ETMC09}.  We do not fit the LECs $\bar\ell_{1,2}$, which
appear at NLO (or $\mathcal{O}(p^4)$) in the finite volume
corrections. They are fixed as in Ref.~\cite{Colangelo:2001df} with their
mass independent parts set to
\begin{equation}
\bar\ell_1^{\,phys} = -0.4 \pm 0.6 \,, \qquad \bar\ell_2^{\,phys} = 4.3 \pm 0.1 \,.
\end{equation}

We have investigated in our fitting, the effects of turning on the pion
mass splitting ($K$) and higher order $\mathcal{O}(a^2)$ discretization
effects ($D_m$ and $D_f$) in various combinations. We have also
investigated the impact of including the $M_{\pi^0}$ data on our fit. We do
not include the $M_{\pi^0}$ data {\it \`a priori} because it is of a much
lower quality compared to the $M_{\pi^\pm}$ data (the uncertainty
associated with it is at least an order of magnitude larger), and it has
rather different systematics. Since the finite volume corrections are 
expected to be $\mathcal{O}(1\%)$, which would be easily subsumed by the 
$\mathcal{O}(10\%)$ error in $M_{\pi^0}$, we do not apply them to the 
$M_{\pi^0}$ data when including them in our fitting.\footnote{We have
  checked explicitly that applying finite volume corrections to the
  $M_{\pi^0}$ data do not alter our fit in any way.} Nevertheless, this 
should be done when the neutral pion mass can be calculated more reliably 
in the future, and we have provided the full NLO resummed formula for FVEs 
associated with $M_{\pi^0}$ in Sec~\ref{sec:RAF}. 

The results of our fits are shown in Table~\ref{tab:sel}.
\begin{table}[tbp]
\begin{ruledtabular}
\begin{tabular}{llllll}
Fit                  & I       & II         & III        & IV          & V* \\
\hline                                                        
$\bar\ell_3$         & 3.44(14)   & 3.31(16)    & 3.45(13)   & 3.35(15)   & 3.38(11)   \\
$\bar\ell_4$         & 4.69(3)    & 4.63(4)     & 4.75(5)    & 4.70(6)    & 4.71(6)    \\
$F_\pi/F$            & 1.0748(7)  & 1.0736(9)   & 1.0760(10) & 1.0749(11) & 1.0752(11) \\
$2B_0\mu_q/M_\pi^2$  & 1.0281(12) & 1.0269(14)  & 1.0283(12) & 1.0273(14) & 1.0276(10) \\
\hline                                                        
$a_C\,(\mathrm{fm})$ & 0.0665(6)  & 0.0676(7)   & 0.0644(22) & 0.0649(28) & 0.0647(24) \\
$a_B\,(\mathrm{fm})$ & 0.0856(6)  & 0.0871(9)   & 0.0811(40) & 0.0815(48) & 0.0813(43) \\
$a_D\,(\textrm{fm})$ & 0.0528(5)  & 0.0536(6)   & 0.0519(12) & 0.0523(15) & 0.0521(14) \\
\hline                                                        
$a_C^4 K$            & --          & 0.0030(4)  &  --        & 0.0032(11) & 0.0028(4)  \\
$a_C^2 D_m$          & --          & --         & 0.055(28)  & 0.055(38)  & 0.057(35)  \\
$a_C^2 D_f$          & --          & --         & 0.038(32)  & 0.045(44)  & 0.044(39)  \\
\hline
$\chi^2/n_\mathrm{dof}$ & 37.5/16   & 32.6/15 & 20.6/14 & 16.1/13 & 21.8/19 \\
$p$-value             & 0.00       & 0.01    & 0.11   &0.24    & 0.29    \\
\end{tabular}
\end{ruledtabular}
\caption{\label{tab:sel} Results of all our fits of the ETMC data from
  ensembles $B_1-B_4,B_6,B_7,C_1-C_3$ and $D_1$~\cite{ETMC09}. The 
  asterisk besides the fit number indicates the inclusion of the
  $M_{\pi^0}$ data. The quantities $M_\pi$ and $F_\pi$ are physical pion
  mass and decay constant, and $\mu_q$ is the value of the quark mass
  corresponding to the physical pion mass.} 
\end{table}
To establish a baseline, we performed Fit~I with pure ChPT fitting forms that 
included neither
twisted mass nor NNLO discretization effects, and we see it gives the worst
description of the data. Turning on just the pion mass splitting in Fit~II 
induced only a marginal improvement, as the $\chi^2$ decreased by 2.4\% with 
the number of degrees of freedom, $n_\mathrm{dof}$, reduced by one. However, 
turning on instead $D_m$ and $D_f$ in Fit~III produced a dramatic improvement
with a 45\% reduction in the $\chi^2$ compared to Fit~I as $n_\mathrm{dof}$
reduced by two. A further improvement is gained when all parameters associated
with discretization effects, $K$, $D_m$ and $D_f$, are simultaneously turned
on in Fit~IV, as the $\chi^2$ decreased another 22\% compared to Fit~III with
$n_\mathrm{dof}$ reduced by one. Finally the best fit is Fit~V* when
$M_{\pi^0}$ data are also included (indicated by the asterisk beside the
fit number).

For each fit, we have also calculated the $p$-value, namely the probability
of obtaining a normalized $\chi^2$ greater than that actually found from
the fit. It is particularly illuminating to compare the $p$-values in
Table~\ref{tab:sel}, as it shows that despite the substantial improvement
in $\chi^2$, Fit~III is not yet fully convincing in terms of its $p$-value.
Only after including the twisted mass effects as in Fit~IV does the
$p$-value rise to a much more acceptable level.  Finally, the inclusion of
the $M_{\pi^0}$ data gives a further increase in the $p$-value in Fit~V*.

\begin{table}[t]
\begin{ruledtabular}
\begin{tabular}{ccc@{$\qquad$}ccc@{$\qquad$}ccc}
\mc{3}{c}{Ensemble data$\quad$} & \mc{3}{c}{Fit~III$\qquad\;\!$} & \mc{3}{c}{Fit~IV$\;\;$} \\
Ensemble & $L/a$ & $a\mu_q$ 
& $\chi^2_{M_\pi}$ & $\chi^2_{F_\pi}$ & $\chi^2_\mathrm{pair}$ 
& $\chi^2_{M_\pi}$ & $\chi^2_{F_\pi}$ & $\chi^2_\mathrm{pair}$ \\
\hline
$B_1$ & 24 & 0.0040 & 3.16 & 0.51 & 3.24 & 0.63 & 0.42 & 2.23 \\
$B_2$ & 24 & 0.0064 & 1.84 & 0.00 & 3.74 & 1.91 & 0.00 & 3.68 \\
$B_3$ & 24 & 0.0085 & 0.03 & 0.39 & 0.40 & 0.00 & 0.20 & 0.23 \\
$B_4$ & 24 & 0.0100 & 1.61 & 0.40 & 1.87 & 1.52 & 0.37 & 1.76 \\
$B_6$ & 32 & 0.0040 & 1.37 & 0.53 & 1.37 & 0.19 & 0.03 & 0.50 \\
$B_7$ & 32 & 0.0030 & 0.02 & 0.57 & 1.64 & 0.74 & 0.01 & 1.87 \\
\hline
$C_1$ & 32 & 0.003  & 4.85 & 1.04 & 4.88 & 3.76 & 0.61 & 3.76 \\
$C_2$ & 32 & 0.006  & 0.16 & 0.06 & 0.38 & 0.18 & 0.02 & 0.30 \\
$C_3$ & 32 & 0.008  & 0.00 & 1.25 & 1.46 & 0.07 & 0.66 & 0.99 \\
\hline
$D_1$ & 48 & 0.0020 & 1.20 & 0.96 & 1.51 & 0.58 & 0.49 & 0.75 \\
\end{tabular}
\end{ruledtabular}
\caption{\label{tab:chidata}
The contribution of each individual ensemble to the total $\chi^2$ 
for Fit~III and IV. The quantities $\chi^2_{M_\pi}$ and $\chi^2_{F_\pi}$ 
denote the contribution each $M_\pi$ and $F_\pi$ data makes separately to the 
total $\chi^2$ neglecting the correlation between them. The full contribution 
from the correlated pair is denoted by $\chi^2_\mathrm{pair}$.}
\end{table}
We remark here that for the pion mass-squared splitting, $a^2 K =
M_{\pi^\pm}^2 - M_{\pi^0}^2 \equiv \Delta M_\pi^2$, its fitted value
changes little whether $M_{\pi^0}$ data are included in the fit or not.
Incidentally, including $M_{\pi^0}$ data seems to reduce the uncertainty in
$a^2 K$ substantially, and leads to a result which is (statistically)
different from zero by about $7\sigma$ compared to a bit under
$3\sigma$ when $M_{\pi^0}$ data are not included. By combining the analysis
of the FVEs of the charged pion masses and the $M_{\pi^0}$ data, one can
see unambiguously the mass splitting predicted in tmChPT, and that $\Delta
M_\pi^2 > 0$.  This represents the first determination of the tmChPT
quantity $W_0^2 W'$ from tmLQCD simulations. We emphasize here that $\Delta
M_\pi^2$, and hence $M_{\pi^0}$, can already be determined from the FVEs
alone. Although large uncertainties are associated with the determination
of $M_{\pi^0}$ either through analyzing FVEs or by direct lattice
calculation, it is reassuring to see that both provide compatible results.

To provide further insight into exactly where improvements arise by taking into account
twisted mass effects, we give a breakdown of the individual contributions to the total 
$\chi^2$ from each ensemble for Fit~III and IV in Table~\ref{tab:chidata}.
We would like to see the individual contribution from each $M_\pi$ and
$F_\pi$ data, and this is given by $\chi^2_{M_\pi}$ and $\chi^2_{F_\pi}$ in
Table~\ref{tab:chidata}.  Note that since $M_\pi$ and $F_\pi$ are
correlated non-trivially in any given ensemble, $\chi^2_{M_\pi}$ and
$\chi^2_{F_\pi}$ do not sum up to the full contribution from that
particular ensemble, $\chi^2_\mathrm{pair}$, which takes into account the
correlation between the pair: they are $\chi^2$ values calculated for each
$M_\pi$ and $F_\pi$ data separately neglecting the correlation

A comparison of Fit~III and IV shows very clearly for which ensemble and for
which quantity is the improvement from taking into account twisted mass 
effects the most significant. The largest improvements are seen in $M_\pi$ for
ensembles $B_1$ and $C_1$, which indeed have small volumes and small pion 
masses. For these two ensembles, the relative improvement in $F_\pi$ is also 
quite evident. In all, we see clear improvements across-the-board.

\section{\label{sec:concl} Conclusions}
In this paper we have analyzed the finite volume corrections to pion masses
and decay constant in tmChPT. The presence of a lighter neutral pion gives
exponentially enhanced finite volume corrections, and we have calculated 
these to NLO accuracy. We have performed a detailed analysis of the ETMC 
data and found out that they are better fitted by the formulae derived here
than by the continuum ones~\cite{CDH05}, particularly in regards to the
finite volume dependence. 
We are able to extract the pion mass splitting predicted in tmChPT from
analyzing FVEs on charged pions alone without having to calculate directly
the neutral pion mass. An important benefit of this is, as far as we know, 
a first determination of the LEC $W'$ of tmChPT. This example shows that, 
though small, FVEs can be successfully used to determine interesting physical 
observables. 

Other LECs of tmChPT appear in parity-violating cubic interactions, which
give exponentially enhanced FVEs. Despite the fact that they are formally of 
higher order in the combined twisted mass chiral expansion, we have calculated 
their analytical form and investigated their effects in our fitting in the hope
that the exponential enhancement may be large enough to allow them to be seen, 
and thus enable extraction of more LECs of tmChPT. Unfortunately, their 
contribution is found to be small, and below the precision of the present ETMC data.

As the precision of the lattice calculations increases and as the
simulations move towards lighter pions, the corrections discussed here will
become even more important and may have an impact also on the extracted
physically relevant parameters. We suggest that future tmLQCD lattice
studies perform analyses of FVEs taking full account of the twisted mass
discretization effects, as was done in deriving our formulae in this paper.

\section*{Acknowledgments}
The authors wish to thank Karl Jansen and Carsten Urbach for discussions
and comments on the manuscript. The Albert Einstein Center for Fundamental
Physics is supported by the ``Innovations- und Kooperationsprojekt C-13''
of the ``Schweizerische Universit\"atskonferenz SUK/CRUS''. Partial
financial support by the Helmholtz Association through
the virtual institute ``Spin and strong QCD'' (VH-VI-231), by the Swiss
National Science Foundation, and by EU MRTN--CT--2006--035482 (FLAVIA{\it
  net}) is gratefully acknowledged. This work has been completed while two
of us (GC and JMSW) were visiting the Galileo Galilei
Institute for Theoretical Physics in Florence. Its hospitality and the
partial support by INFN are gratefully acknowledged.

\end{document}